\documentclass[preprints,article,accept,moreauthors,pdftex]{Definitions/mdpi} 
\usepackage{caption}
\usepackage{float}
\firstpage{1} 
\pubvolume{xx}
\issuenum{1}
\articlenumber{5}
\pubyear{2019}
\copyrightyear{2019}
\history{Received: 04 November 2019; Accepted: 27 December 2019; Published: date}

\pdfoutput=1

\Title{Influence of Resonances on the Noise Performance of SQUID Susceptometers}

\Author{{Samantha I. Davis} $^{1,*, \dagger,\S}$, {John R. Kirtley} $^{2,\dagger}$ and { Kathryn A. Moler} $^{1,2,3,4}$}

\AuthorNames{Samantha I. Davis, John R. Kirtley and Kathryn A. Moler}

\address{%
$^{1}$ \quad Department of Physics, Stanford University, Stanford, CA 94305-4045, USA; s1dav1s@alumni.stanford.edu \\
$^{2}$ \quad Geballe Laboratory for Advanced Materials, Stanford University, Stanford, CA 94305, USA;\\
$^{3}$ \quad Department of Applied Physics, Stanford University, Stanford, CA 94305, USA; kmoler@stanford.edu\\
$^{4}$ \quad Stanford Institute for Materials and Energy Sciences, SLAC National Accelerator Laboratory, 2575 Sand Hill Road, Menlo Park, CA 94025, USA; kmoler@stanford.edu (K.A.M.)}

\corres{Correspondence: {s1dav1s@alumni.stanford.edu}}

\firstnote{These authors contributed equally to this work.}
\thirdnote{{Current address: California Institute of Technology -- The Division of Physics, Mathematics and Astronomy
1200 E California Blvd, Pasadena CA 91125, USA}} 

\abstract{Scanning Superconducting Quantum Interference Device (SQUID) Susceptometry simultaneously images the local magnetic fields and susceptibilities above a sample with sub-micron spatial resolution. Further development of this technique requires a thorough understanding of the current, voltage, and flux ($IV\Phi$) characteristics of scanning SQUID susceptometers. These sensors often have striking  anomalies in their current--voltage characteristics, which we believe to be due to electromagnetic resonances. The effect of these resonances on the performance of these SQUIDs is unknown. To explore the origin and impact of the resonances, we develop a model that qualitatively reproduces the experimentally-determined $IV\Phi$ characteristics of our scanning SQUID susceptometers. We use this model to calculate the noise characteristics of SQUIDs of different designs. We find that the calculated ultimate flux noise is better in susceptometers with damping resistors that diminish the resonances than in susceptometers without damping resistors. Such calculations will enable the optimization of the signal-to-noise characteristics of scanning SQUID susceptometers.}


\begin{document}


\section{Introduction}
 
\begin{figure}[H]
    \centering
    \includegraphics[width=0.8\textwidth, trim=0 0 0 0]{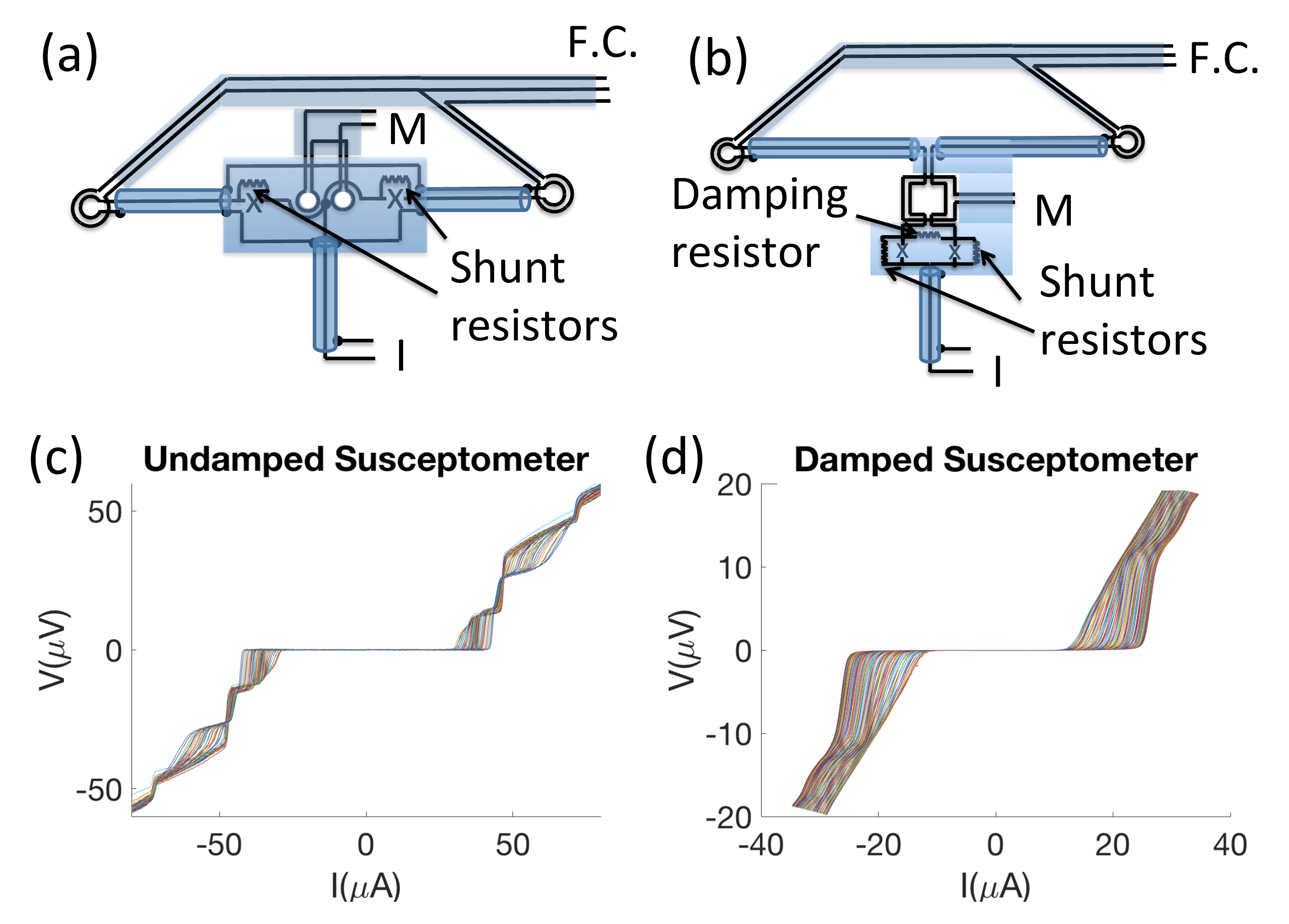}
   \caption{{{ Two types of susceptometer layouts}:} ({\bf{a}}) That of Huber et al. \cite{huber2008gradiometric,Kirtley2016}, without a damping resistor, and {(\bf{b}}) that of Gardner et al., \cite{gardner2001scanning} with a damping resistor. $I$ labels the current leads, $M$ the modulation coil leads, and $F.C.$ the field coil leads. The Josephson junctions are indicated by $X$s. The semi-transparent regions indicate superconducting shields. Superconducting coaxial leads connect the central regions with junctions and modulation coils to the pickup loop/field coil pairs to the left and right. ({\bf{c}}) Current--voltage (IV) characteristic for an undamped susceptometer at various magnetic fluxes, and ({\bf{d}}) IVs for a damped susceptometer. }
  \label{fig:two_layouts}
\end{figure}

Superconducting Quantum Interference Devices (SQUIDs) are superconducting loops interrupted by one or more Josephson weak links \cite{clarke2004squid}. SQUIDs are used to achieve high-precision magnetic sensing for diverse applications, including gravitational-wave astrophysics \cite{vinante2001dc,harry2000two}, magnetoencephalography \cite{hari2012magnetoencephalography}, quantum information \cite{devoret2004superconducting}, and scanning SQUID microscopy \cite{kirtley1999scanning}. In scanning SQUID microscopy (SSM), SQUIDs are used to image the local magnetic fields above samples. Enhanced spatial resolution is achieved in SSM by either making very small SQUID loops \cite{black1993magnetic,veauvy2002scanning,finkler2010self} or integrating a small ``pickup loop" into the body of a larger SQUID through well-shielded superconducting coaxial leads \cite{vu1993imaging,kirtley1995high}. An extension of SSM is scanning SQUID susceptometry, in which susceptibility measurements are made by surrounding the pickup loop in the latter type of SQUID by a co-planar, co-axial single-turn field coil \cite{gardner2001scanning}, often in a gradiometric configuration (see Figure \ref{fig:two_layouts}). As sensitive techniques for probing mesoscopic materials, scanning SQUID magnetometry and susceptometry are paving the way for essential advances in superconductor physics \cite{kirtley2010fundamental}. 

One necessary step to advancing scanning SQUID technologies is understanding scanning SQUID behaviors. In this paper, we analyze the current--voltage--flux ($IV\Phi$) properties of scanning SQUID susceptometers. Typically, sensitive SQUID magnetic flux measurements are made using a flux-locked loop \cite{clarke2004squid}. Here, we calculate the behaviors of our SQUID susceptometers when current-biased: The voltage across the SQUID at constant current is held fixed by feeding back on the flux through the SQUID using a modulation coil (see Figure \ref{fig:two_layouts}a,b). The flux through the modulation coil compensates for changes in the flux through the pickup loop. If there is sufficient feedback through the flux-locked loop, the current through the modulation coil is proportional to the flux through the pickup loop. The sensitivity of the SQUID is due to the fact that near the critical current, small changes in flux result in large changes in voltage. 

Scanning SQUID susceptometers also measure magnetic susceptibility by applying a localized magnetic field to the sample through the field coils. SQUID susceptometers are laid out in a gradiometric configuration so that they are insensitive to both uniform magnetic fields and currents that pass through both field coils (Figure \ref{fig:two_layouts}a,b). With this layout, SQUID susceptometers image the local magnetic flux and magnetic susceptibility of materials directly below one of the pickup loop/field coil pairs simultaneously.

To optimize the performance of scanning SQUID sensors, it is important to understand their $IV\Phi$ characteristics in the presence of noise. Although the noise properties of ideal SQUIDs are well understood \cite{Tesche1977,bruines1982comment}, the noise properties of new-generation SQUIDs, such as SQUID susceptometers, are not.
One puzzling phenomenon is the presence of anomalies in the $IV\Phi$ characteristics of SQUID susceptometers (Figure \ref{fig:two_layouts}c). These anomalies take the form of ``steps" in voltage (peaks in the dynamic resistance $dV/dI$), which occur at currents that disperse strongly with applied magnetic flux $\Phi$ and have a period of one superconducting flux quantum ($\Phi_0$). Previous studies have shown that the current--voltage and {alternating current (a.c.)} 
characteristics of SQUIDs can be affected by parasitic capacitances in their input circuitry \cite{hilbert1985measurements,enpuku1992modeling,huber2001dc} and that the SQUIDs' performances can be improved by resistive damping of the resultant input coil resonances~\cite{knuutila1987effects}. We believe our anomalies are of similar origin and therefore refer to them as ``resonances." 

To explore the origin and impact of resonances on the $IV\Phi$ characteristics of SQUID susceptometers, we perform an analysis of {two-junction, direct current (d.c.)} 
 SQUID susceptometers that seeks the answers to two queries: 1) What causes the resonances? and 2) Do the resonances enhance or diminish the sensitivity of SQUID susceptometers? 

To address the first query, we develop models of susceptometers that reproduce the resonances in simulations of their $IV\Phi$ characteristics. We hypothesize that the resonances occur due to parasitic capacitances and inductances that arise from complex features of the SQUIDs, such as the field coils, the gradiometric layout, and pickup loops that are integrated into the bodies of the SQUIDs through superconducting coaxial leads (Figure \ref{fig:two_layouts}). These parasitic inductances and capacitances introduce {inductor-capacitor (LC)} 
 resonances that are driven by the a.c. Josephson oscillations of the junctions in the voltage state. Consequently, when the LC resonance frequency matches the Josephson frequency, there are voltage steps in the IV$\Phi$ characteristics, which translate to peaks in the IR$\Phi$ characteristics of the susceptometers. 


Our hypothesis is supported by basic estimates of the voltage steps. The resonances in our susceptometers have a characteristic voltage of roughly 10 $\mu$V (see Figure \ref{fig:two_layouts}c). Combining the Josephson relations \cite{josephson1962possible}, 
\begin{eqnarray}
    I_s&=& I_0 \sin \varphi \nonumber \\
    V&=&\frac{1}{2\pi} \frac{d\varphi}{dt},
    \label{eq:josephson}
\end{eqnarray}
where $I_s$ is the supercurrent through the junction, $I_0$ is the junction critical current, $V$ is the voltage, and $\varphi$ is the quantum mechanical phase drop across the junction, with the resonance frequency for an LC circuit,  
\begin{equation} \label{eq:LC}
    \omega = \frac{1}{\sqrt{L_{p,eq}C_{p,eq}}},
\end{equation}
 we expect voltage steps to occur at
\begin{equation} \label{eq:VLC}
    V_{LC}=\frac{\Phi_0}{2\pi \sqrt{L_{p,eq}C_{p,eq}}},
\end{equation}
where $L_{p,eq}$ and $C_{p,eq}$ are the equivalent lumped parasitic inductance and capacitance of the circuit, respectively. To estimate $L_{p,eq}$ and $C_{p,eq}$, we use the FASTCAP, FASTHENRY, and INDUCT software packages from the Whiteley Research web site \cite{WRSPICE}. We estimate that the susceptometer of Figure \ref{fig:two_layouts}a has $L_{p,eq} = 60\pm20$ pH and $C_{p,eq} = 20\pm 6$ pF, which results in a characteristic voltage of $6.9~\mu V < V_{LC} < 12.7~\mu V$. Using this intuition, we are able to successfully simulate the complex resonant behavior of the susceptometers. We also reproduce the behavior of susceptometers with damping resistors, which greatly reduce the amplitude of the resonances. 

In what follows, we first demonstrate in Section \ref{sec:IRPHI} that the addition of a parasitic capacitance to the standard model for a SQUID produces peaks in the $IR\Phi$ characteristics similar to those observed experimentally (see Figure \ref{fig:w_wout_R}). We then show that we can qualitatively reproduce the highly complex $IR\Phi$ characteristics of an undamped SQUID using a relatively simple model with distributed parasitic inductances and capacitances, as well as the much simpler $IR\Phi$ characteristics that result when a damping resistor is introduced (see Figure \ref{fig:two_ladder}). We proceed to calculate SQUID noise in Section \ref{sec:noise}, first demonstrating that we can reproduce previous work on basic SQUID layouts. After confirming our procedure, we calculate the noise in our more complicated undamped and damped models at selected positions in the $IR\Phi$ plane. We conclude that the lowest intrinsic noise in the damped layout is significantly lower than that in the undamped layout for susceptometers for parameters that give similar critical curves.

\section{Modeling}

\subsection{IR$\Phi$ Characteristics}
\label{sec:IRPHI}
We use commercial software to model our devices: XIC, a layout tool, and WRSPICE, a simulation tool, both developed by Whiteley Research \cite{WRSPICE}. WRSPICE is based on the JSPICE \cite{JSPICE} simulation tool for electronic circuits and includes Josephson junctions. The layout tool XIC produces a list of nodes that specify connections between devices from a schematic. In our case, the devices are resistors $R$, capacitors $C$, inductors $L$, mutual inductances $M$, and Josephson junctions $JJ$. Each device has a constitutive equation: $V=IR$ for the resistors, $Q=CV$ for the capacitors, $V=LdI/dt$ for the inductors, and the Josephson relations (Equation {\ref{eq:josephson}) for the Josephson junctions.

The nodes, devices, and constitutive equations are combined by WRSPICE into a matrix equation of the form $AX=B$, where the elements of the vector $X$ are the device responses and the elements of $B$ are the excitations (e.g., voltage and current sources). In general, the matrix equation is non-linear and is solved by LU (lower, upper) decomposition iteratively with Newton's method. In our case, we do a transient analysis that produces the time dependence of the circuit response in the presence of d.c. biases, magnetic flux, and noise. 

We assume that the pairs of critical currents $I_0$ and shunt resistances $R_J$ for the two Josephson junctions are identical for each SQUID. To calculate the IV characteristics at each flux $\Phi$, we ramp the current through the SQUID at a rate of 1~$\mu$A/nsec and average the resulting voltage time trace (which has large Josephson oscillations) in bin widths of 1 $\mu$A. Figure \ref{fig:w_wout_R}b displays typical results for an ``ideal" SQUID with parasitic inductance $L_p$ but no parasitic capacitance $C_p$.

 \begin{figure}[H]
    \centering
    \includegraphics[width=0.8\textwidth, trim=0 0 0 0]{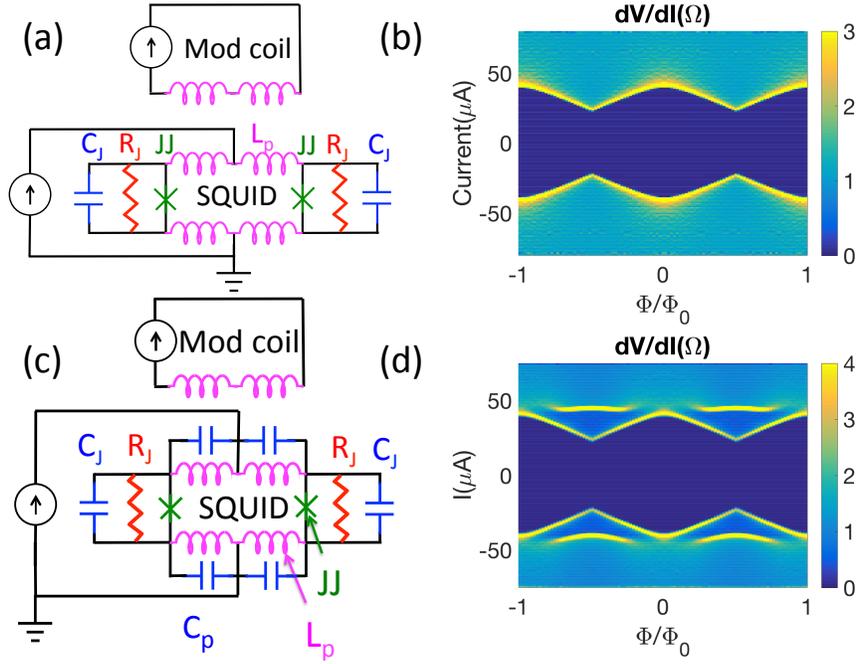}
   \caption{{{Adding a parasitic capacitance to an ideal Superconducting Quantum Interference Device (SQUID) produces a resonance.}} ({\bf{a}}) Ideal SQUID schematic, and ({\bf{b}}) calculated dV/dI characteristic for an ideal SQUID with no parasitic capacitance at $T$ = 4.2 K. In this instance, the upper inductances $L_p$ = 30 pH, the lower inductances $L_p$ = 1 pH, the Josephson critical currents $I_0$ = 22 $\mu A$, the shunt resistors $R_J$ = 2 $\Omega$, and the junction capacitances $C_J$ = 10 fF. ({\bf{c}}) Schematic with a parasitic capacitance, and ({\bf{d}}) calculated dV/dI characteristic at $T$ = 4.2 K. Here $I_0$ = 22 $\mu$A, $R_s$ = 2 $\Omega$, $C_j$ = 10 fF, upper $L_p$=30 pH, lower $L_p$ = 1 pH, upper $C_p$=10 pF, and lower $C_p$ = 1 pF.}
  \label{fig:w_wout_R}
\end{figure}

  Figure \ref{fig:w_wout_R}c displays the schematic and Figure \ref{fig:w_wout_R}d displays the dV/dI characteristic for the same circuit as in Figure \ref{fig:w_wout_R}a, but with parasitic capacitances $C_p$ added in parallel with the parasitic inductances. These capacitances could result from, e.g., the overlapping superconducting layers between the junctions in Figure \ref{fig:two_layouts}a. In this case, there are single resonances at half-integer multiples of $\Phi_0$ and voltages of $\approx 19~\mu V$ ($\approx 42~\mu A$), but no resonances at higher voltages. The resonances occur at junction voltages in good agreement with Equation (\ref{eq:VLC}), taking $L_{p,eq}=30~pH$ and $C_{p,eq}=10~pH$. The simulations also show strong peaks in the variance of the current through the parasitic inductors at 19~$\mu$V, supporting the hypothesis that the resonances arise when the Josephson oscillations drive the parasitic $LC$s at their resonance frequency.

The more complicated schematic of Figure \ref{fig:two_ladder}a qualitatively reproduces the complex behavior of the resonances seen experimentally for an undamped susceptometer (see Figure \ref{fig:two_ladder}b,c). In this case, the resonances are generated in a ``ladder" of paired $L_p$s and $C_p$s, which physically correspond to the distributed inductances and capacitances of the superconducting coaxes leading to the pickup loops. We find that there is not a one-to-one correspondence between the number of resonances and the number of $L_pC_p$ pairs, but rather that the fine details of the resonances depend on the number and values of $L_pC_p$ pairs included in the simulation. The details of the model (listed in the caption of Figure \ref{fig:two_ladder}) are chosen to fit the experiment by tweaking the various parameters. The quality of the fit is measured by calculating the mean variance between the model and calculated $IR\Phi$ characteristic $\chi^2 = \sum_{n,m} (R_{\rm exp.}(I_n,\Phi_m) - R_{\rm model}(I_n,\Phi_m))^2/N$, where $N$ is the total number of calculated points in the $I\Phi$ plane. The model results displayed in Figure \ref{fig:two_ladder}c correspond to $\chi^2=2.91~\Omega^2$. We are not able to find a set of $I_0$, $C_p$, $L_m$, and $L_p$ parameters that cause the modeled peaks in $IR\Phi$ to perfectly overlap with the experimental peaks. Nevertheless, we find the qualitative agreement between experiment and modeling exhibited in Figure \ref{fig:two_ladder}b,c supports the hypothesis that the structure in the IV characteristics is due to $LC$ resonances driven by Josephson oscillations.

\begin{figure}[H]
    \centering
    \includegraphics[width=1.0\textwidth, trim=0 0 0 0]{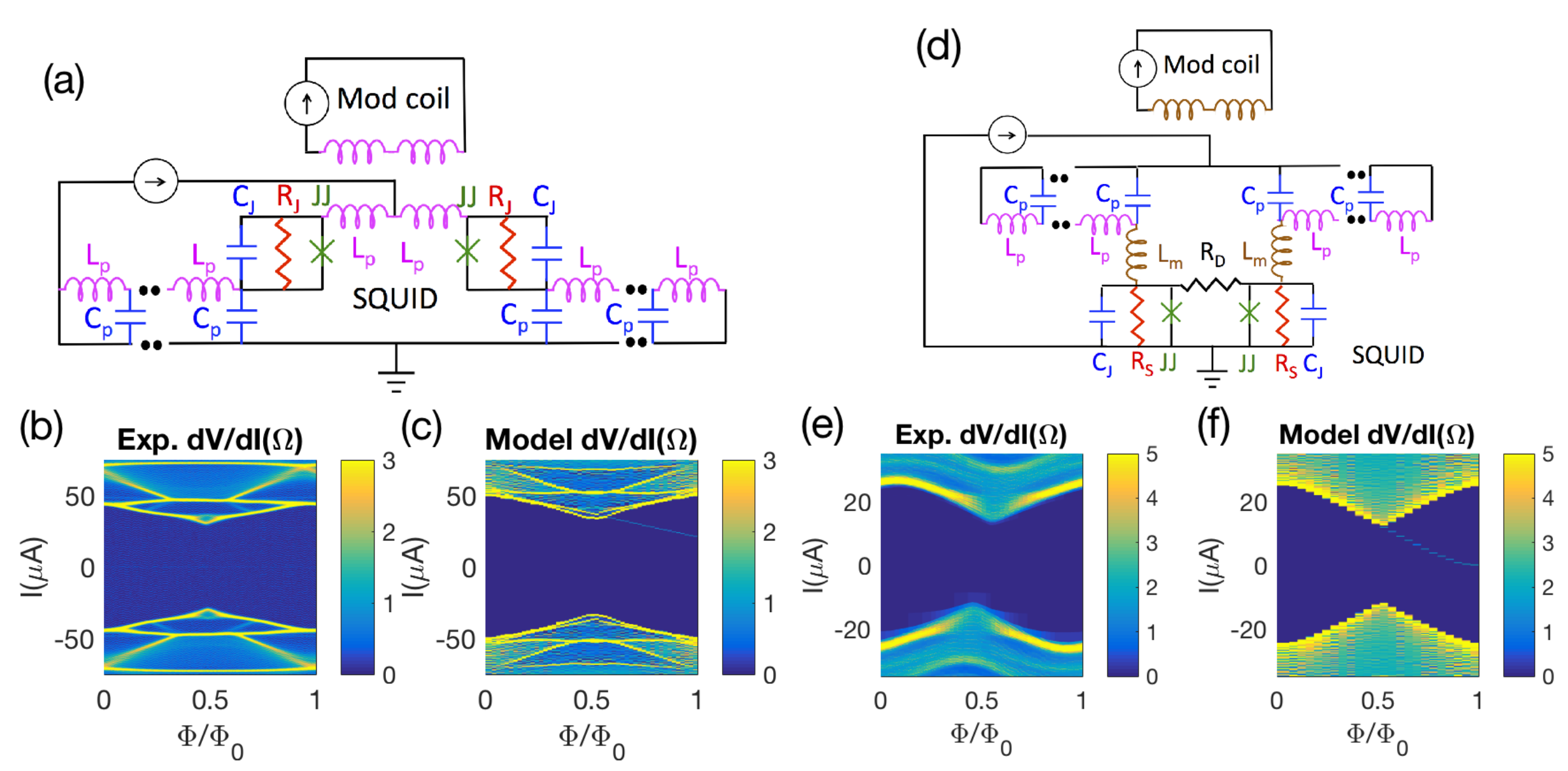}
   \caption{{{Modeling of dV/dI vs. I }and $\Phi$ ($IR\Phi$) for two types of  SQUID susceptometers:} ({\bf{a}}) Undamped schematic, ({\bf{b}}) experimental dV/dI characteristic, and ({\bf{c}}) calculated dV/dI characteristic at $T$ = 4.2 K for a SQUID with the layout of Figure \ref{fig:two_layouts}a \cite{huber2008gradiometric,Kirtley2016}. In this model, $I_0$ = 25 $\mu$A, $R_J$ = 2~$\Omega$, $C_J$ = 10 fF, $L_m$ = 30 pH, $L_p$ = 4 pH, and $C_p$ = 8 pF. There are a total of five $L_p, C_p$ pairs in each arm to the left and right of the schematic, representing the coaxial leads to the pickup loops. ({\bf{d}}) Damped schematic, ({\bf{e}}) experimental dV/dI characteristic, and ({\bf{f}}) calculated dV/dI  characteristic at $T$ = 4.2 K for a SQUID with the layout of Figure \ref{fig:two_layouts}b \cite{gardner2001scanning}. In this model, $I_0$ = 12 $\mu$A, $R_J$ = 2~$\Omega$, $R_D$ = 4 $\Omega$, $C_j$ = 10fF, $L_m$ = 30 pH, $L_p$ = 1 pH, and $C_p$ = 8 pF. There are five $L_p, C_p$ pairs in each arm of the center of the schematic, representing the coaxial leads to the pickup loops. }
  \label{fig:two_ladder}
\end{figure}

We also find that in both the modeling and experiment, the resonances can be greatly reduced with the addition of a damping resistor. Figure \ref{fig:two_ladder}d displays the schematic of a SQUID susceptometer with a damping resistor (see Figure \ref{fig:two_layouts}b, \cite{gardner2001scanning}), with parameters adjusted to fit the experimental IV$\Phi$ characteristics.

\begin{table*}[ht]
\caption{{Dimensionless parameters.}} 
\centering 
\begin{tabular}{|c | c | c |}
\hline                        
Parameter & Symbol &  Conversion formula  \\
 [0.5ex]
\hline  
Voltage & $v$ & $V/I_0R_J$    \\
Magnetic flux & $\phi$   & $\Phi/\Phi_0 $ \\
Thermal noise parameter   & $\Gamma$ & $2\pi k_bT/I_0\Phi_0$  \\
Voltage noise power   & $S_v^0$ & $2\pi S_V^0/I_0R_J\Phi_0 $   \\
Flux noise & $\zeta_\phi^{1/2}$ & $S_{\Phi}^{1/2}(\pi I_0R_J/\Gamma)^{1/2}/\Phi_0^{3/2}$ \\
Hysteresis parameter & $\beta$ & $2 LI_0/\Phi_0$ \\
[1ex] 
\hline 
\end{tabular}
\label{table:tesche_units} 
\end{table*}

\subsection{Noise}
\label{sec:noise}
 The characteristic time step for the transient analysis in JSPICE is a fraction of the inverse Josephson frequency---typically several GHz. Since we are interested in the noise at frequencies of several hundred Hz or below, such calculations can be very time consuming (see the discussion in Ref. \cite{Tesche1977}). Noise is introduced into our simulations as Johnson noise from the resistors with a Gaussian distributed voltage in series with the resistors with standard deviation $V_n=\sqrt{2 k_b T R/dt}$, or equivalently, current sources in parallel with the resistors with standard deviation $I_n=\sqrt{2 k_b T/dtR}$, where dt is the time interval. For the noise results we report here, we fix the flux and current through the SQUID and solve for the voltage as a function of time, typically recording the voltage $V(t)$ in 1 ps intervals over 300 ns. We then Fourier transform $V(t)$ to get the power spectral density $S_V(f)$, fit the results below the frequency $\langle V(t)\rangle/10~\Phi_0$ (where $\langle V(t)\rangle$ is the average voltage over the full time trace) to a straight line, and extrapolate to zero frequency to obtain $S_V^0$. The data from any currents that have fewer than 100 points in this frequency interval or have a negative intercept from the linear fit are rejected. The transfer function $dV/d\Phi$ is obtained by subtracting two runs separated by 0.02~$\Phi_0$ centered on the flux of interest, and the flux noise is $\sqrt{S_\Phi^0}$= $\sqrt{S_V^0}/(dV/d\Phi)$. We repeat this procedure ten times. 
 Following Tesche and Clarke \cite{Tesche1977}, we report our results using reduced units. Table \ref{table:tesche_units} lists these units and conversion formulas to obtain them from S.I. units. In this table, $k_b$ is Boltzman's constant, $\Phi_0=h/2e$ is the superconducting flux quantum, $I_0$ is the single junction critical current, $R_J$ is the single junction shunt resistance, and $T$ is the temperature.

We first verify that we can reproduce previous work. Figure \ref{fig:Tesche}a displays the schematic, and Figure~\ref{fig:Tesche}b--d displays the dimensionless voltage noise power $S_v^0/2\Gamma$, the dimensionless transfer function $|dv/d\phi|$, and the dimensionless flux noise $\zeta_{\phi}^{1/2}$ respectively for the models used by Tesche and Clarke \cite{Tesche1977} and Bruines et al. \cite{bruines1982comment}. The results have several qualitative features that are common to all the models studied: The dynamic resistance $dV/dI$ (not shown), the voltage noise power $S_v^0$ (Figure \ref{fig:Tesche}b), and the transfer function $|dv/d\phi|$ (Figure \ref{fig:Tesche}c) have peaks, and the flux noise $\zeta_{\phi}^{1/2}$ (Figure \ref{fig:Tesche}d) has a broad minimum, at similar flux -ependent currents $I$. The error bars for $S_v^0/2\Gamma$, $dv/d\phi$, and $\zeta_\phi^{1/2}$ are calculated through error propagation. Using the standard error propagation formula, we find that the statistical uncertainty for $S_v^0/2\Gamma$ is proportional to the standard deviation of the fit to the voltage periodogram at low frequencies. Since the amplitude of the voltage noise is greater at large $dV/dI$, the statistical uncertainty in $S_v^0/2\Gamma$ is greater in the vicinity of the critical curve and resonance. For $dv/d\phi$, we find that the uncertainty in $dv/d\phi$ is proportional to the sum in quadrature of the uncertainties of the voltages at the two fluxes, so the error bars are also greater near the critical curve and resonance. For $\zeta_\phi ^{1/2}= V_{noise}/(dv/d\phi)$, the contribution from the error of $dv/d\phi$ is proportional to $(dv/d\phi)^{-1}$, so the uncertainty in $\zeta_\Phi ^{1/2}$ is large where $dv/d\phi$ is close to zero. The results from Bruines et al. \cite{bruines1982comment} differ from those by Tesche and Clarke \cite{Tesche1977} in the transfer function $|dv/d\phi|$ and flux noise $\zeta_{\phi}^{1/2}$ because of a numerical error in Tesche and Clarke. Our results agree with the results of Bruines et al. \cite{bruines1982comment} to within statistical uncertainty.

\begin{figure}[H]
   \centering
   \includegraphics[width=0.5\textwidth, trim=0 0 0 0]{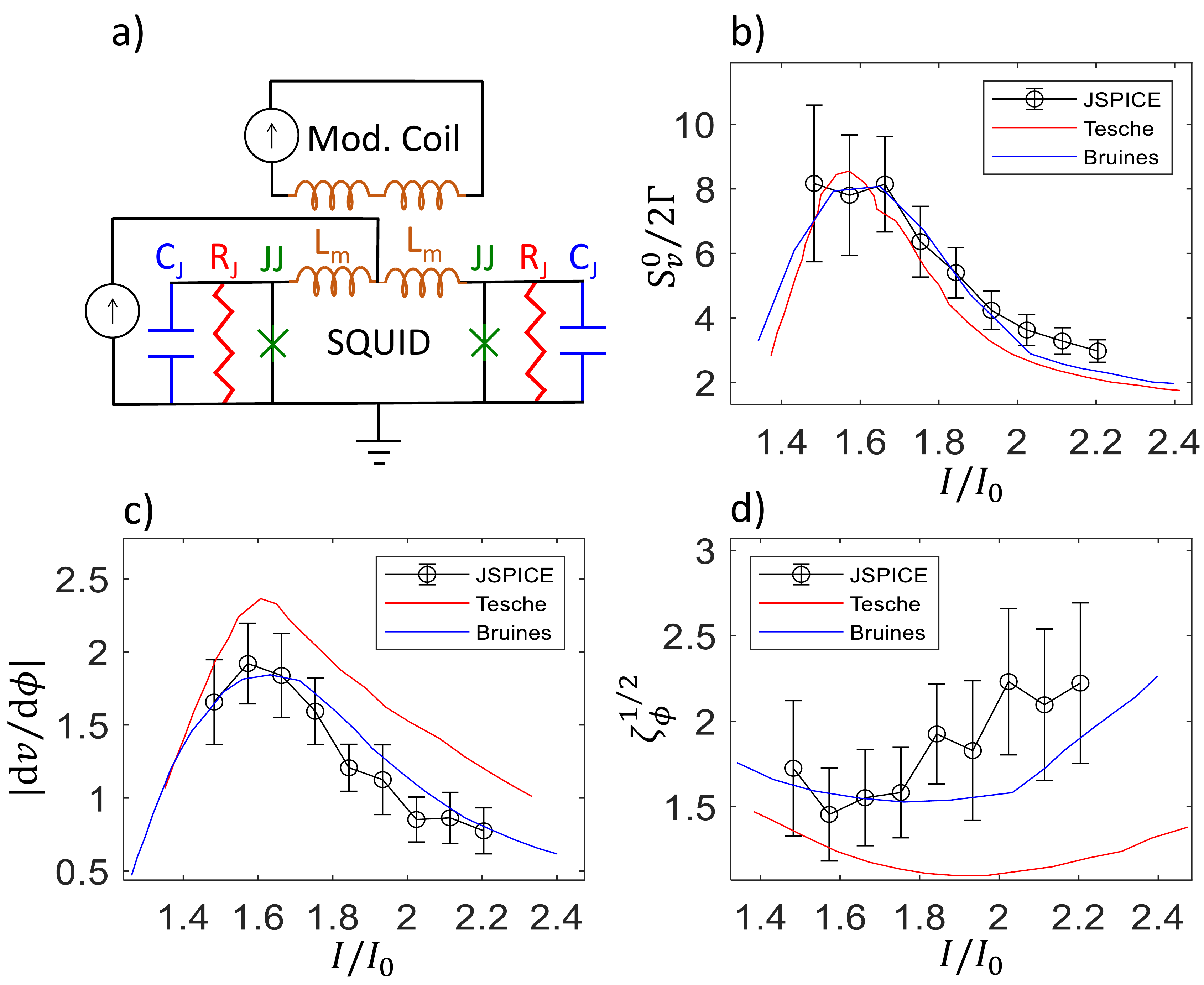}
    \caption{{{ Comparison with previous work}:} ({\bf{a}}) Schematic of the model used. In this case, the junction critical current $I_0$ = 17.2~$\mu$A, junction capacitance $C_j$ = 0 pF, modulation inductance $L_m$ = 30 pH, shunt resistance $R_J$ = 2 $\Omega$, $\Phi = 0.25~\Phi_0$, and $T$ = 20.56 K. This choice of parameters leads to $\beta = 1.0$, $\Gamma = 0.05$, for direct comparison with Figures 13a, 14a, and 15a of Tesche and Clarke \cite{Tesche1977}, as well as Figures 1a and 2a of Bruines et al. \cite{bruines1982comment}. The curve labelled Bruines in ({\bf{b}}) is inferred from the curves labelled Bruines in ({\bf{c}}) and ({\bf{d}}).}
    \label{fig:Tesche}
\end{figure}

After confirming our procedure by reproducing previous work, we then proceed to calculate the noise for the more complicated undamped susceptometer, with conceptual layout given by Figure~\ref{fig:two_layouts}a, and schematic given by Figure \ref{fig:two_ladder}a. It would take prohibitively long to calculate the noise for all currents and fluxes. Instead, we choose four values for flux at currents along the ``critical curve", at which the junction is just entering the voltage state. The symbols superimposed on the $dV/dI$ plot in Figure~\ref{fig:undamped_CC}a show the values of current $I$ and flux $\phi$ used for each calculation, with paired fluxes (required to calculate $|dv/d\phi|$) separated by 0.02 $\Phi_0$ and centered on each flux value plotted in Figure \ref{fig:undamped_CC} b--d. 
\begin{figure}[H]
\centering
\includegraphics[width=0.5\textwidth, trim=0 0 0 0]{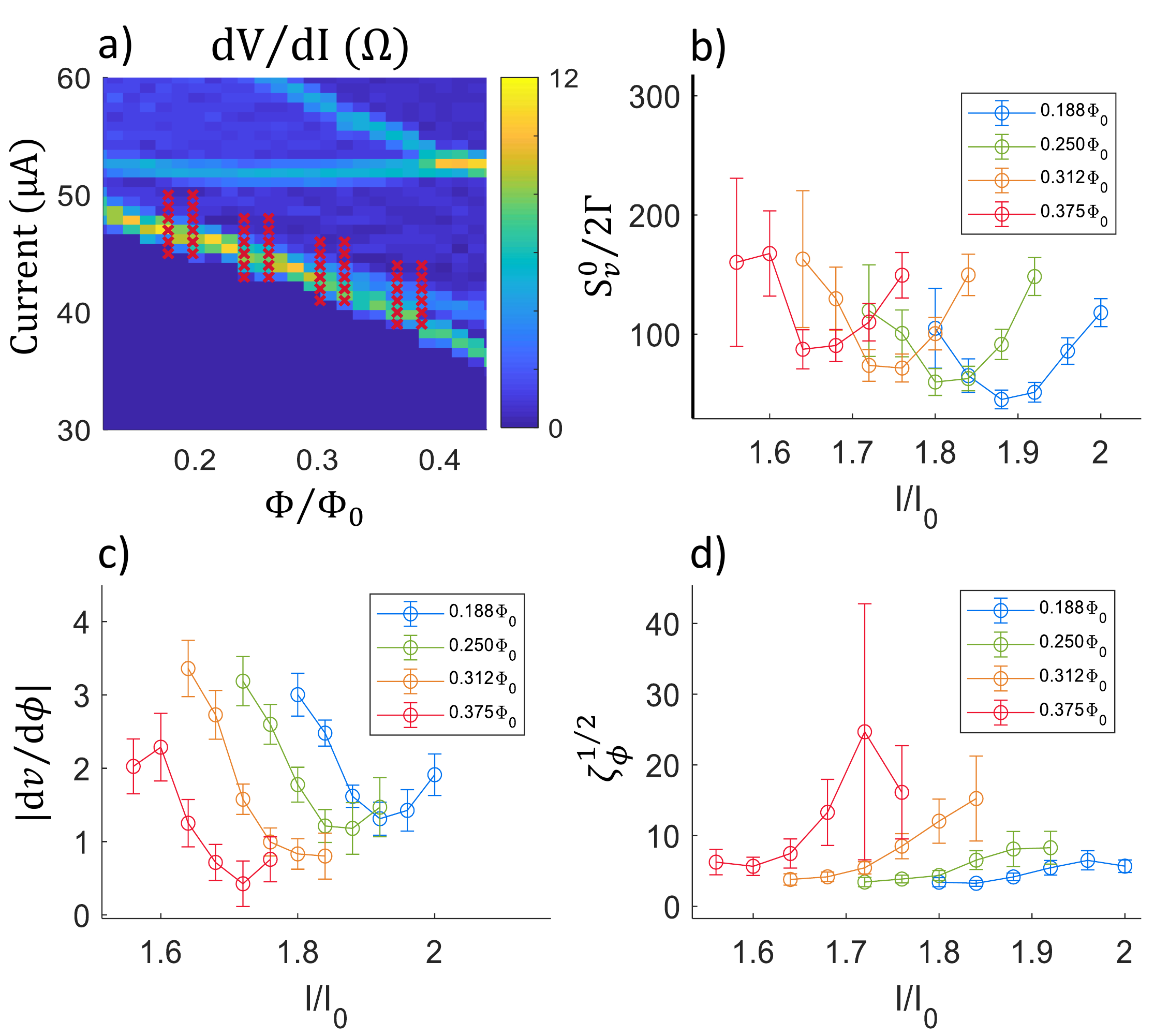}
\caption{{{Noise calculations for an undamped SQUID along the critical curve}:} ({\bf{a}}) Plot of dV/dI vs. current (I) and flux ($\Phi$). The crosses correspond to the values of $I$ and $\Phi$ for which noise was calculated. There are two sets of crosses, separated by 0.02 $\Phi_0$, to enable the calculation of the derivative $dv/d\phi$ at each flux value. ({\bf{b}}) Plots of the dimensionless low-frequency voltage noise power $S_v^0/2\Gamma$ vs. current $I$ for four different flux values. ({\bf{c}}) Plots of the dimensionless transfer junction $|d\nu/d\phi|$. ({\bf{d}}) Plots of the dimensionless flux noise $\zeta_{\phi}^{1/2}$. The schematic used for these calculations was that of Figure \ref{fig:two_ladder}a, with $I_0$ = 25 $\mu$A, $L_m$ = 30 pH, $R_J$ = 2 $\Omega$, $L_p$ = 4 pH, $C_J$ = 10 fF, $C_p$ = 8 pF, and $T = 4.2$ K.
}
\label{fig:undamped_CC}
\end{figure}

Figure \ref{fig:undamped_R} displays similar calculations for the same model and parameters as Figure \ref{fig:undamped_CC}, but for current and flux values along the first resonance in the IR$\Phi$ characteristic.

\begin{figure}[H]
\centering
\includegraphics[width=0.5\textwidth, trim=0 0 0 0]{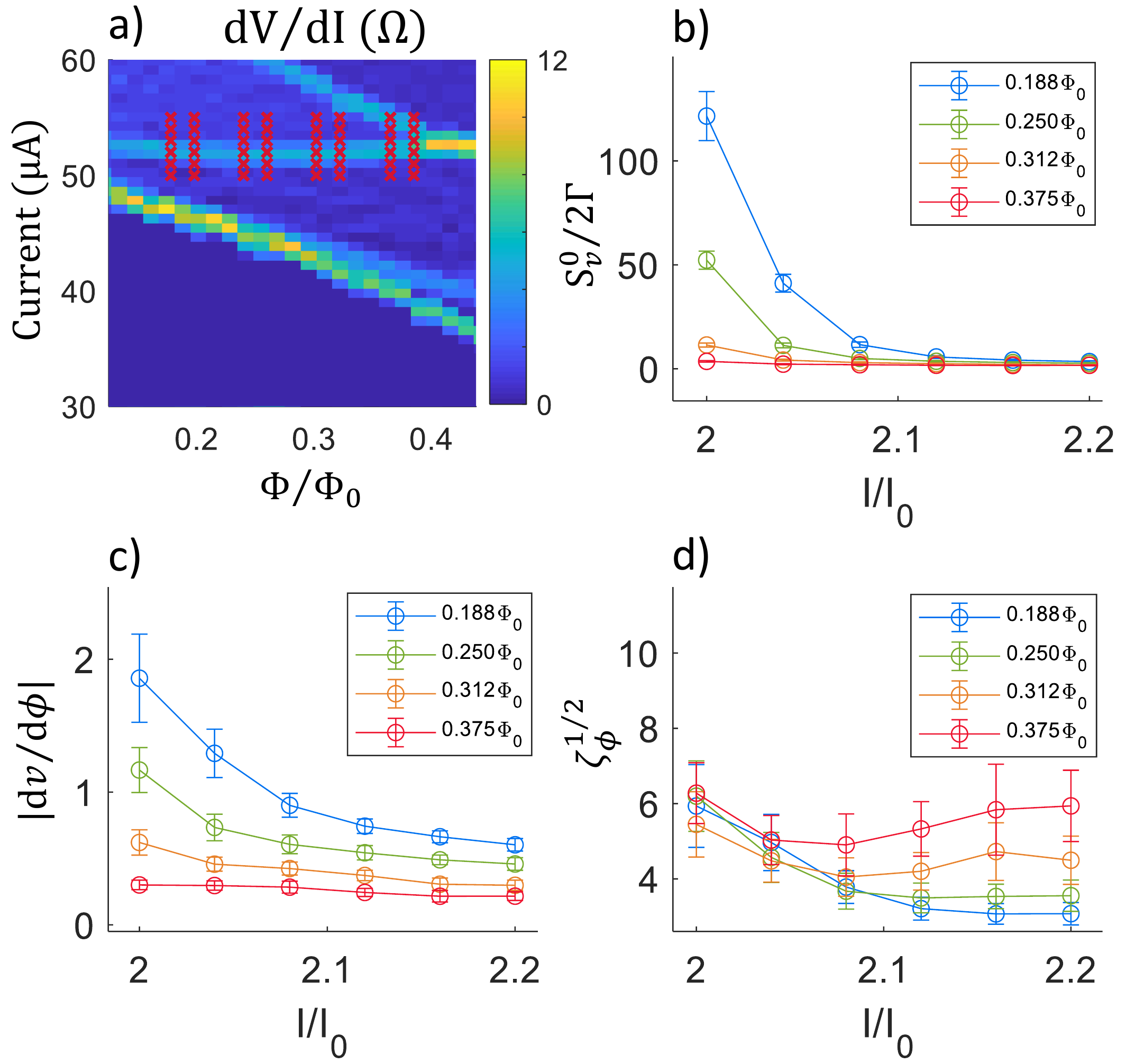}
\caption{{{Noise calculations for an undamped SQUID along resonance at the lowest magnitude  bias current} ("first resonance"):} ({\bf{a}}) Plot of dV/dI vs. current (I) and flux ($\Phi$). The crosses correspond to the values of $I$ and $\Phi$ for which noise was calculated. ({\bf{b}}) Plots of the dimensionless low-frequency voltage noise power $S_v^0/2\Gamma$ vs. current $I$ for four different flux values. ({\bf{c}}) Plots of the dimensionless transfer junction $|d\nu/d\phi|$. ({\bf{d}}) Plots of the dimensionless flux noise $\zeta_{\phi}^{1/2}$. The schematic used for these calculations was that of Figure \ref{fig:two_ladder}a, with $I_0$~= 25 $\mu$A, $L_m$ = 30 pH, $R_J$= 2 $\Omega$, $L_p$ = 4 pH, $C_J$ = 10 fF, $C_p$ = 8 pF, and $T$ = 4.2 K.
}
\label{fig:undamped_R}
\end{figure}

Finally, Figure \ref{fig:noise_damped} displays results for the damped SQUID susceptometer model with layout in Figure \ref{fig:two_layouts}b and schematic in Figure \ref{fig:two_ladder}d. For these calculations, the parameters were chosen to match those for the undamped SQUID, except for the addition of a damping resistor $R_d = 4~\Omega$, for more direct comparison between the damped and undamped cases.

\begin{figure}[H]
   \centering
   \includegraphics[width=0.5\textwidth, trim=0 0 0 0]{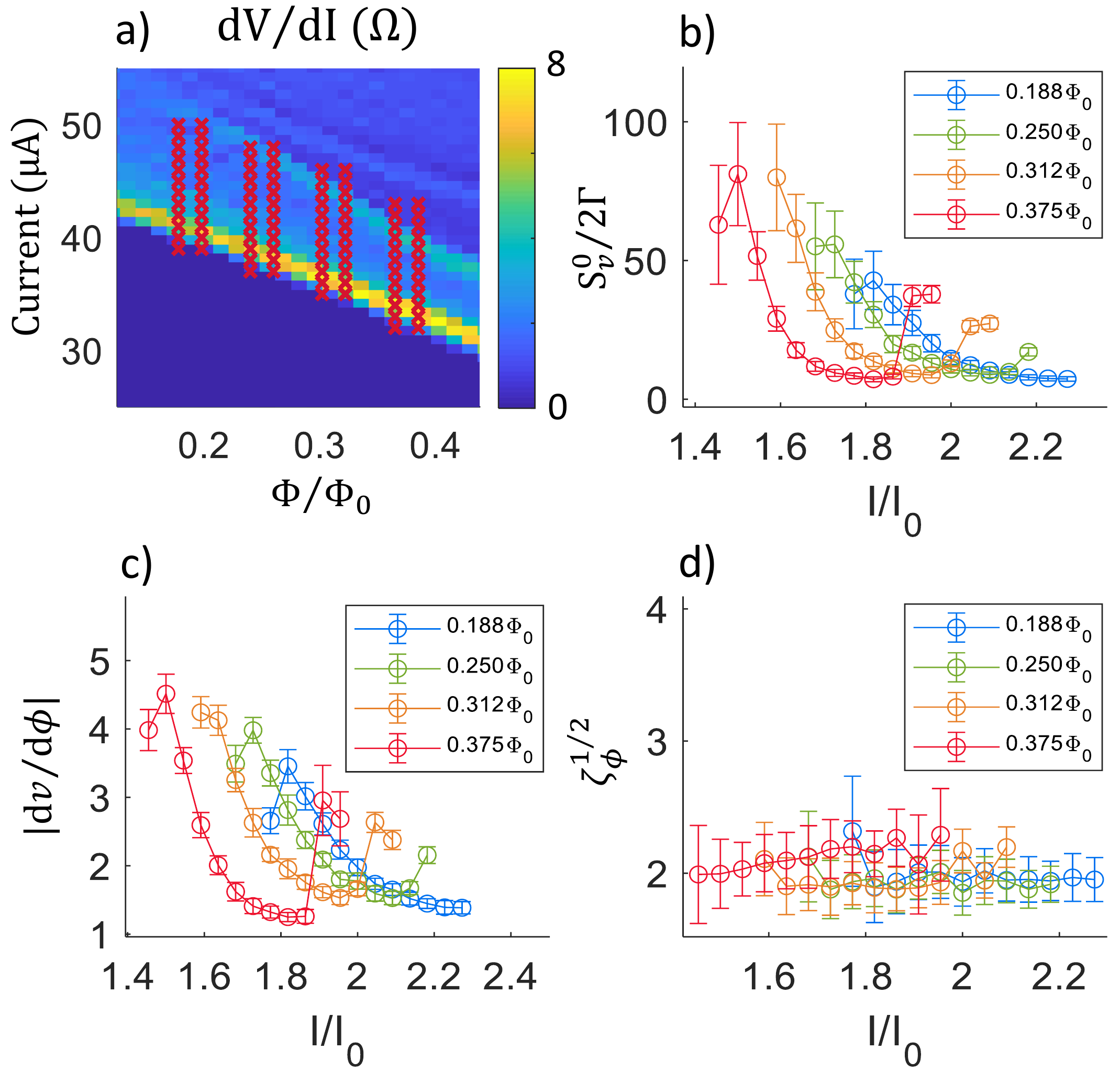}
    \caption{{{Noise calculations for a damped SQUID along the critical curve:}} ({\bf{a}}) Plot of dV/dI vs. current (I) and flux ($\Phi$). The crosses correspond to the values of $I$ and $\Phi$ for which noise was calculated. ({\bf{b}}) Plots of the dimensionless low-frequency voltage noise power $S_v^0/2\Gamma$ vs. current $I$ for four different flux values. ({\bf{c}}) Plots of the dimensionless transfer junction $|d\nu/d\phi|$. ({\bf{d}}) Plots of the dimensionless flux noise $\zeta_{\phi}^{1/2}$. The schematic used for these calculations was that of Figure \ref{fig:two_ladder}d, with $I_0$ = 22~$\mu$A, $L_m$ = 30 pH, $R_J$ = 2 $\Omega$, $R_d$ = 2 $\Omega$, $L_p$ = 1 pH, $C_J$ = 10 fF, $C_p$ = 8 pF, and $T$ = 4.2 K.}
    \label{fig:noise_damped}
\end{figure}

\subsection{Summary of Noise Calculations}
A summary of the noise analysis described in Section \ref{sec:noise} is reported in Figure \ref{fig:MinFluxNoisevsFlux_3pts}. 
The calculated minimum flux noise is similar for the undamped susceptometer model on the critical curve vs. on the first resonance, but is significantly lower for the damped susceptometer model than for the undamped one for parameters that give similar critical curves. 

\begin{figure}[H]
\centering
\includegraphics[width=0.5\textwidth, trim=0 0 0 0]{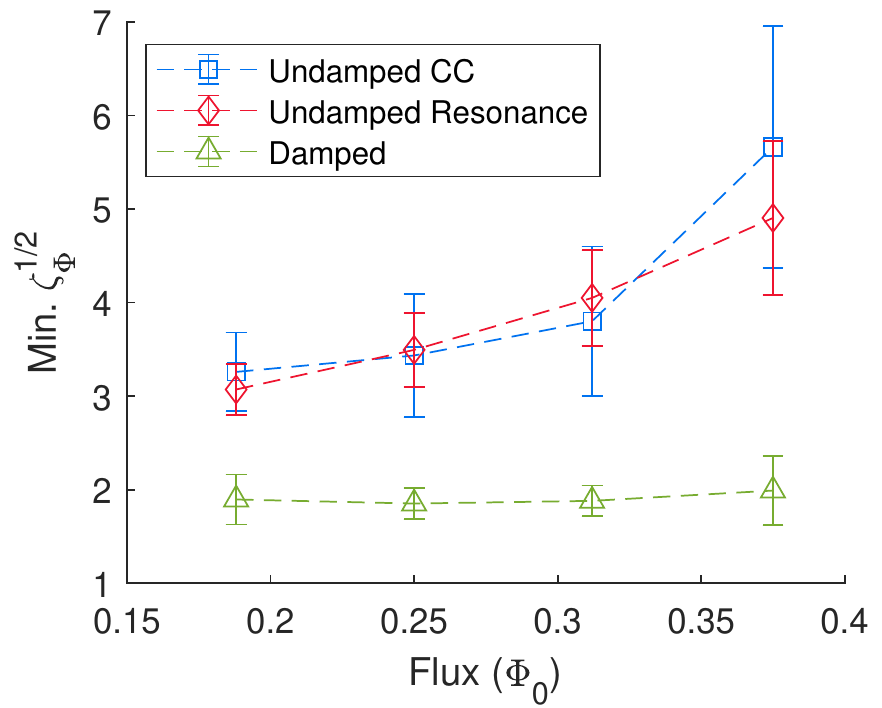}
\caption{{{Minimum flux noise for damped vs. undamped SQUIDs.}} The square blue symbols correspond to the undamped SQUID along the critical curve, the diamond red symbols are for the undamped SQUID along the first resonance, and the triangular green symbols correspond to the damped SQUID along the critical curve. These data were generated by varying the current ($I$) at fixed flux. 
}
\label{fig:MinFluxNoisevsFlux_3pts}
\end{figure}

The noise values that we calculate are comparable to experimentally reported noise floors for scanning SQUID susceptometers. The undamped SQUIDs presented in Figure \ref{fig:two_layouts}c have a critical current $I_0$= 25~$\mu A$ and a shunt resistance $R_J$=2 $\Omega$.  At $T = 4.2$ K, a reduced flux noise of $\zeta_{\phi}^{1/2}=3$ corresponds to $S_{\phi}^{1/2}=0.91~\mu\Phi_0/Hz^{1/2}$ for these values of $I_0$ and $R_J$. The damped SQUIDs presented in Figure \ref{fig:two_layouts}d have a critical current $I_0$=12.5~$\mu A$ and shunt resistance of $R_J$=4~$\Omega$, so $\zeta_{\phi}^{1/2}=2$ corresponds to $S_{\phi}^{1/2}=0.86~\mu\Phi_0/Hz^{1/2}$ at 4.2~K. Gardner et al. \cite{gardner2001scanning} report an intrinsic noise of 3~$\mu\Phi_0/Hz^{1/2}$ for damped scanning SQUID susceptometers at 4.2~K. Kirtley et al. \cite{Kirtley2016} report an intrinsic noise of 2~$\mu\Phi_0/Hz^{1/2}$ for undamped susceptometers at 4.2~K, while Huber et al. \cite{huber2008gradiometric} report a noise of 0.25~$\mu\Phi_0/Hz^{1/2}$ at 125~mK above 10~kHz for undamped susceptometers.

\authorcontributions{{Conceptualization}, J.R.K.; Formal analysis, S.I.D. and J.R.K.; Funding acquisition, K.A.M.; Software, S.I.D. and J.R.K.; Writing – original draft, S.I.D. and J.R.K.; Writing – review and editing, S.I.D. and J.R.K.}

\funding{{This research received no external funding.}}

\acknowledgments{We would like to thank Diana Chamaki for making some of the IV measurements of damped susceptometers presented in this paper.}

\conflictsofinterest{{The authors declare no conflict of interest.}}

\reftitle{References}

\end{document}